\documentclass[lettersize,journal]{IEEEtran}
\usepackage{amsmath,amsfonts}
\usepackage{algorithmic}
\usepackage{algorithm}
\usepackage{array}
\usepackage{textcomp}
\usepackage{url}
\usepackage{verbatim}
\usepackage{graphicx}
\usepackage{cite}
\usepackage{subcaption}
\begin{document}

\title{LLMs are everywhere: Ubiquitous Utilization of AI Models through Air Computing}
\author{Baris Yamansavascilar, Atay Ozgovde, and Cem Ersoy}

\maketitle

\begin{abstract}

We are witnessing a new era where problem-solving and cognitive tasks are being increasingly delegated to Large Language Models (LLMs) across diverse domains, ranging from code generation to holiday planning. This trend also creates a demand for the ubiquitous execution of LLM-powered applications in a wide variety of environments in which traditional terrestrial 2D networking infrastructures may prove insufficient. A promising solution in this context is to extend edge computing into a 3D setting to include aerial platforms organized in multiple layers, a paradigm we refer to as air computing, to augment local devices for running LLM and Generative AI (GenAI) applications. This approach alleviates the strain on existing infrastructure while enhancing service efficiency by offloading computational tasks to the corresponding air units such as UAVs. Furthermore, the coordinated deployment of various air units can significantly improve the Quality of Experience (QoE) by ensuring seamless, adaptive, and resilient task execution. In this study, we investigate the synergy between LLM-based applications and air computing, exploring their potential across various use cases. Additionally, we present a disaster response case study demonstrating how the collaborative utilization of LLMs and air computing can significantly improve outcomes in critical situations.

\end{abstract}


\section{Introduction}
\IEEEPARstart{L}{arge} Language Models (LLMs) are currently superstars of artificial intelligence (AI) since their easy utilization and responsiveness have provided many benefits for both companies and end-users. To this end, they can be utilized for many different tasks including chat systems, image processing/generation, data prediction, and content creation. Therefore, ChatGPT, DALL-E, and LLaMA can be given as examples for popular LLM applications.

LLMs are a result of long-term studies starting from statistical language models (SLMs) such as n-gram to pre-trained attention-based transformer models such as BERT and GPT \cite{zhou2024comprehensive}. On the other hand, the context-aware word representations through word embeddings in pre-trained models accelerated the evolution of natural language processing (NLP) methods \cite{mikolov2013distributed, vaswani2017attention}. Hence, by providing the pre-trained models along with the transformer architecture by using the bigger model and data size for the training, the capacity of LLMs is quite improved so that they can show emergent abilities, including reasoning and instruction-following. As a result of this, essential LLM models, which are called foundation models, can be configured to more specific tasks to adapt to different use cases. Furthermore, their capabilities can allow us to use them for multimodal tasks, such as for a given text input, they can produce an image or video. Thus, LLM models can be used for different goals and therefore can be deployed in different environments. Based on the existing requirements and the corresponding infrastructure, LLMs can dramatically enhance user quality of experience (QoE).

\begin{figure}[t]
\centering
\includegraphics[scale=0.09]{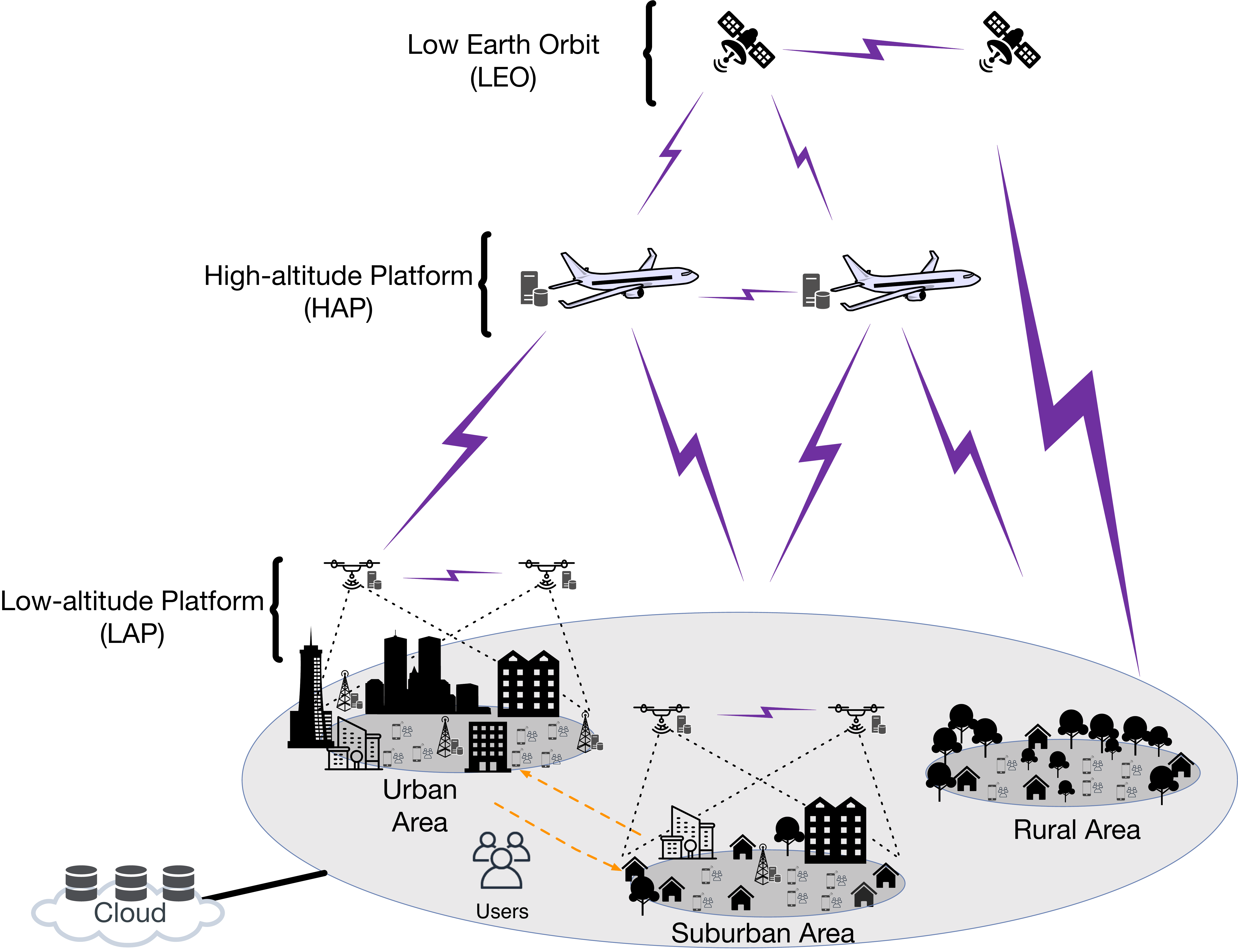}
\caption{An air Computing Environment}
\label{AirComputing-Mag}
\end{figure}

On the other hand, considering networking paradigms and corresponding task offloading/processing strategies, multi-access edge computing (MEC) has provided significant benefits for different application types and user profiles over the years. However, when the infrastructure is not sufficient, its capabilities would not meet the next-generation application requirements and would be insufficient for dynamic capacity provision. Therefore, 3D networking structure through different air layers including low altitude platform (LAP), high altitude platform (HAP), and low earth orbit (LEO) has recently gained attention \cite{mu2023uav, yamansavascilar2024air}. To this end, the coordinated utilization of those layers through air computing including different units such as unmanned aerial vehicles (UAVs), airplanes, and satellites can enhance the quality of service (QoS) of the corresponding applications and therefore improve user quality of experience (QoE). An example of an air computing environment is depicted in Figure \ref{AirComputing-Mag}.

\begin{figure*}[t]
\centering
\includegraphics[scale=0.0255]{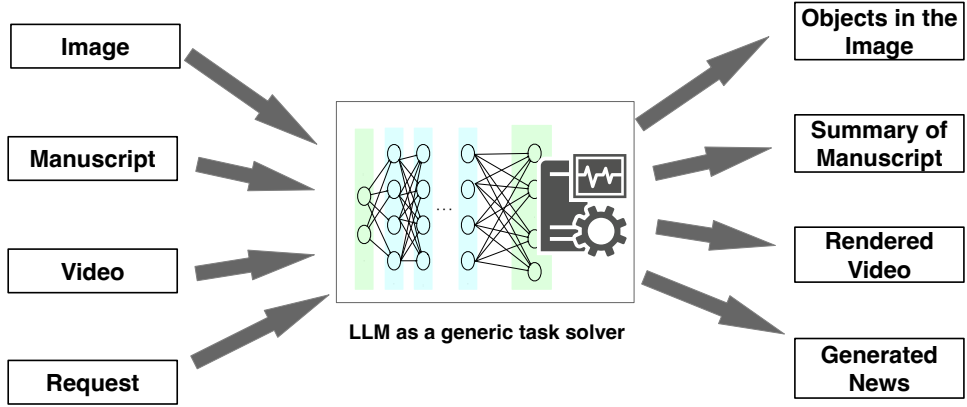}
\caption{LLMs have generalization abilities so that they can process various task types with high performance.}
\label{LLMs-generalization}
\end{figure*}

One of the most crucial features of LLMs is their ability to solve general-purpose tasks as shown in Figure \ref{LLMs-generalization}. Therefore, various types of tasks based on different application profiles can be processed by general-purpose LLMs. For example, an image processing task which can detect an accident scene would be significant for secure transportation. Likewise, detecting a disaster or a possible landslide earlier through different inputs would be crucial for public safety. Moreover, content creation and evaluation would also be useful and alleviate people's daily tasks.



With the wide adoption of Large Language Models (LLMs) and Generative AI (GenAI) applications, finding efficient methods to run these models at scale poses significant computational and networking challenges. Firstly, the user demand may extend beyond the capabilities of conventional urban infrastructure, especially in scenarios where demand occurs in suburban or rural areas. One potential use case is large-scale outdoor events with crowded clusters of participants needing to access the services in remote areas. Another use case covers the emergency response scenarios in disaster areas where existing infrastructure may become completely unusable and survivors needing to access the services. Even in cases where the overall capacity of the infrastructure is sufficient, as in urban scenarios, the capacity offered by air units can help regulating the spikes in computational demand, such as during concerts, festivals, or large public gatherings. In this context, air computing emerges as an effective approach, enabling dynamic augmentation of terrestrial infrastructure through the deployment of various air units \cite{yamansavascilar2024air, zeng2016wireless}.

Using air computing devices in edge computing scenarios necessitates the existence of edge AI units that can be transported in the air. In this context, we are witnessing rapid growth in the capabilities of commercially available AI accelerator card families, such as the NVIDIA Jetson and Google Coral. These devices enable high-performance deep learning inference at the edge while maintaining a lightweight form factor suitable for UAV transportation. Empirical research has already demonstrated the potential of these edge devices to run deep learning models efficiently while achieving low power consumption \cite{swaminathan2024benchmarking}. Complemented by the novel LLM performance optimization techniques like quantization, it is evident that we will increasingly observe the deployment of LLMs on such platforms.

Considering the benefits of air computing and the profits of generic task-solving capabilities of LLMs, a coordinated utilization of them together would bring about advantages regarding application performance and user satisfaction. Therefore, in this study, we investigate those advantages considering different use cases. The main contributions of our study are as follows:

\begin{itemize}

\item We represent the ubiquitous utilization of LLMs for different task types through air computing. Therefore, we think that it can open new horizons for the next-generation applications that leverage LLMs for their benefits.

\item We demonstrate the benefits of air computing for LLM tasks considering QoS and QoE. Hence, we detail several scenarios for different environments that can affect user satisfaction if air units are not deployed properly.

\end{itemize}

The rest of the paper is organized as follows. In Section II, we summarize the LLMs and air computing, respectively. We elaborate on three use cases in Section III providing detailed scenarios. In Section IV, we provide experimental results for a disaster scenario. Finally, in Section V, we conclude our study.













\section{Background}

In this section, we highlight the significant features of both LLMs and air computing which may alleviate to meet the stringent requirements of the next-generation applications if they can be utilized together \cite{cui2024llmind}.

\subsection{Essential LLM Features}
The most prominent characteristic of LLMs is their emergent abilities which can provide higher performance in many different tasks considering earlier language models \cite{wei2022emergent}. Therefore, they can solve complex problems that earlier language models could not solve such as conversation abilities, generalization, in-context learning, and reasoning. The main reason behind those emergent abilities is related to their big scale in terms of the token size and model size, which include hundreds of billions of parameters \cite{kaplan2020scaling}. We investigate three major features brought about by emergent abilities.



\subsubsection{Reasoning}
LLMs can perform reasoning, especially on complex and staged tasks such as coding and mathematical problems. Moreover, if they need additional information for those stages, they use a prompting mechanism and ask users for the related input. Especially, the technique called chain of thought (CoT) allows us to provide corresponding input in detail with examples so that the LLM agent can give a more structured response. Therefore, they outperform earlier language models in the corresponding tasks that require reasoning.

\subsubsection{In-context Learning}
Through the in-context learning feature, LLMs can learn directly from demonstrations and instructions without requiring extra training operation \cite{dong2024survey, floridi2020gpt}. Afterwards, they can apply those learnings to different tasks succesfully. Therefore, LLMs can achieve high success rates on the corresponding tasks including question-answering, and translation.

\subsubsection{Generalization Ability}
One of the most distinct characteristics of LLMs is their task-solving capacities, which can be generalizable for many different task profiles \cite{zhao2023survey}. Considering earlier SLMs, they could be focused on specific tasks only. Moreover, neural and pre-trained language models can learn context-aware representations through word embeddings. However, since they would not manifest emergent abilities due to their small model and token sizes regarding LLMs, those models cannot perform well for generic tasks. On the other hand, LLMs can be used for different goals such as image processing, decision making, text/image/video generating, and context summary. As a result of their generic task-solving capabilities, they can be deployed in different systems, including edge servers, UAVs, and cloud servers, to which different types of tasks are offloaded to be processed, in order to enhance QoE.

The generalization ability of LLMs is ensured through foundation models. These pre-trained models can further be optimized/trained through fine-tuning considering specialized use cases. For example, an LLM model can focus only on image processing or video rendering. Moreover, by using advanced techniques such as Retrieval-Augmented Generation (RAG), external sources can be utilized to increase the reliability of the model. Thus, the accuracy of the corresponding result can better meet the expected outcome.

\subsection{Air Computing}

Air computing is an evolution of edge computing into a 3D networking structure through air layers including LAP, HAP, and LEO \cite{yamansavascilar2024air}. The coordination between different air units and therefore air layers can alleviate the burden on computational resources considering task offloading, content caching, coverage, mobility, computational capacity, and mobility. Since each air layer has distinct features, collaboration between them would be crucial for large-scale deployments considering QoS and QoE. As a result, the main advantage of air computing is that it can meet the dynamically changing requirements of the next-generation applications as well as user needs for various environments. Moreover, air computing can reach a vast amount of users since it can operate in urban, suburban, and rural areas.

\begin{figure*}[t]
\centering
\includegraphics[scale=0.0655]{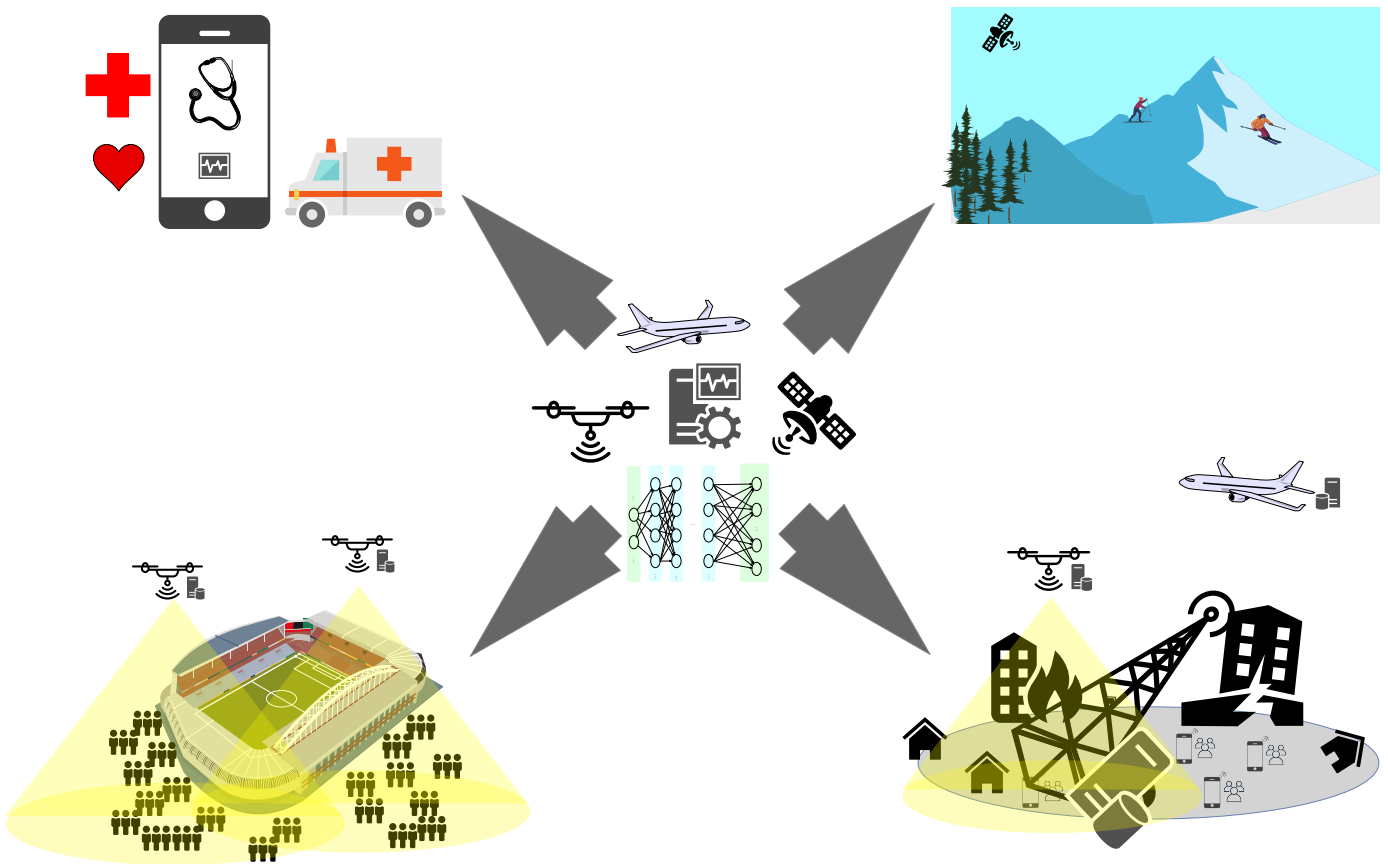}
\caption{The collaborative utilization of LLMs and air computing can provide innovative solutions for various use cases.}
\label{LLM-and-AirComputing}
\end{figure*}

\subsubsection{LAP}
operates between 0 to 10 km, therefore the propagation delay would change between 0 - 30 $\mu$s. Because of its operational altitude, UAVs configured with computational capacities are mainly deployed in LAP as main air vehicles.  Since UAVs require charging stations due to their limited battery capacities, LAP generally operates in urban areas. Moreover, due to low propagation delay, mission-critical applications can be run in LAP-supported environments.

UAVs can use their computational capacity efficiently since they can easily move to locations where user density is high. Moreover, thanks to their easily configurable stationary positions, they can be utilized flexibly. Therefore, they can provide Line of Sight (LoS) communication, which provides seamless connectivity, service provision, and low latency. Apart from their computational assistance for air computing, they can also be used as relay nodes in order to provide seamless mobility.

\subsubsection{HAP}

operates between 10 - 30 km which results in propagation delays between 30 - 100 $\mu$s. Therefore, air vehicles including airplanes and balloons deployed in HAP can be utilized for both urban and suburban areas based on the needs. However, since weather conditions can affect the communication quality in HAP, mission-critical applications cannot be provided well regarding service level agreement requirements. As a result, HAP is fundamentally used for regional coverage where the corresponding air vehicles can be used as controller nodes for UAVs and terrestrial servers.


\subsubsection{LEO}

operates between 160 - 2000 km therefore it consists of satellites. Since the propagation delay is between 0.5 - 7 ms due to high altitude, LEO is not suitable for low-latency applications. Moreover, based on the same practice, LEO is generally utilized for rural areas whose access to computing resources is limited. Hence, they can be used for task offloading considering task processing through their onboard capacity, or as a relay node that conveys those tasks to a nearby air vehicle in an air computing environment. Furthermore, they can also be utilized to access cloud computing resources for heavy tasks such as video rendering.


\section{Use Cases}

Since LLMs have excellent capabilities for task processing/solving through instruction tuning, they can be deployed at terrestrial edge servers and air vehicles in different air platforms. Therefore, the collaborative utilization of LLMs and air computing would provide many opportunities including outdoor activities, remote health, and sports/concert activities as shown in Figure \ref{LLM-and-AirComputing}.


\subsection{Outdoor Activities}

Outdoor activities such as hiking, sailing, climbing, and trekking are generally performed in rural areas which have limited or no infrastructure. Task offloading is difficult to perform as there would be no edge servers in the environment. Furthermore, the services may be more critical in the case of an emergencies in the secluded areas including mountains, forests, and seas.

The corresponding LLM models can be deployed in HAP vehicles, and LEO satellites as a generic task solver, and users can offload their tasks to those units from secluded areas. Hence, air computing and LLM collaboration can enhance user QoE. However, since HAP and LEO would not be sufficient for low-latency applications because of the propagation delay, this solution would be more suitable for non-critical tasks.

\subsection{Remote Health}
People suffering from injuries or chronic diseases may require continuous monitoring or real-time intervention based on the urgency of the corresponding cause. LLMs can assess the existing situation by evaluating the related videos or recent images. Based on the result, the responsible people can be notified about the recent condition of patients.

Even though these cases can be performed well in urban areas through wireless networks or traditional MEC solutions, they would not be carried out appropriately in suburban and rural areas. Considering suburban areas, the edge servers would be limited, and the capacity of home networks can be insufficient. On the other hand, rural areas may not have the necessary infrastructure for the corresponding solutions. To this end, air computing solutions can be deployed in these areas. UAVs deployed with LLM models, which are specialized for monitoring, can be utilized in suburban areas. Likewise, airplanes and balloons can be used in urban areas along with the LLM models. Thus, the coordinated utilization of LLMs and air computing can enhance the recovery of suffered people.

\subsection{Sports and Concert Activities}

Although the infrastructure in an urban area is sufficient most of the time, it may suffer due to an intense load for sports and concert activities. Since there would be many users in the crowded location, the existing infrastructure including edge servers and other terrestrial resources would not meet the required capacity for high load. Moreover, if there are multiple events in the same district or city, the situation would be exacerbated for QoS. Hence, an innovative solution should be applied to this dynamic problem.

Since many of those activities can be performed in urban or suburban areas, UAVs deployed with LLM models can be used to meet the related load in those particular locations. Based on the required capacity, which is computed from the existing load, a sufficient number of UAVs can be sent to the user-concentrated areas. As a result, the capacity can be enhanced dynamically. Thus, users can offload their various tasks to the generic LLM model. As a result, their QoE can be improved significantly. 

\begin{figure*}[t]
\centering
\includegraphics[scale=0.3]{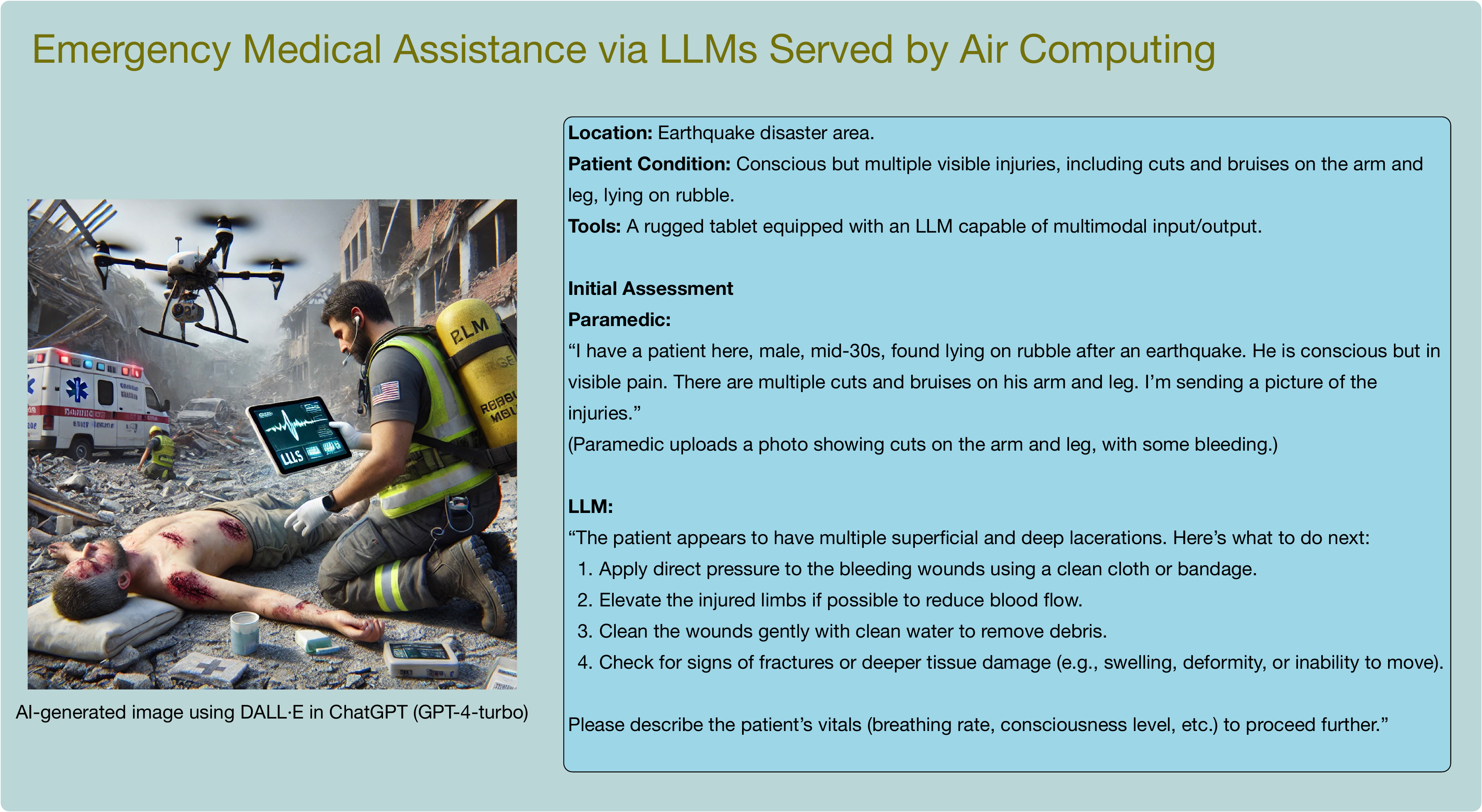}
\label{LLM-Disaster}
\end{figure*}

\subsection{Natural Disasters}

A natural disaster can destroy critical buildings and houses, causing a crucial threat to people's lives. Based on the severity of the disaster, the corresponding infrastructure would also be destroyed. Since people need to contact nearby hospitals and authorities, the loss of communication may start a panic, and people who need medical care can suffer from their injuries. 

In the case of a disaster, the collaborative utilization of air computing and LLMs can provide many advantages such as enhanced early damage assessment through image rendering and, therefore, can identify the destruction level in the environment. Hence, authorities can list the relevant units, vehicles, and required tools in advance and dispatch them to the disaster area. On the other hand, another advantage is that through air vehicles, communication can continue between disaster locations and non-affected areas. In this case, LLMs can also be used to list the corresponding actions to people who may be affected by the disaster at different levels, and report these to authorities.


\section{Case Study: A Disaster Scenario}
We selected the disaster case to test the performance of air computing deployed with LLMs through UAVs in the case of an earthquake. We evaluated the performance based on the task success rate of the corresponding towns affected by the earthquake at different levels.

\subsection{Scenario}

In our scenario, we have three towns which are affected by the disaster at different extents. Town-1 is the most affected area as its infrastructure is completely destroyed. On the other hand, Town-2 severely senses the earthquake with mediocre damage. Finally, Town-3 senses the earthquake mildly.

Each town normally operates an edge server that provides low-latency AI-powered services for its users. However, following the earthquake, the destruction of the infrastructure combined with the increasing demand for computational resources due to critical emergency-related activities results in service outages. The influx of distressed users needing information and assistance and the limited availability of medical personnel in disaster areas further amplifies the situation by leaving many injured individuals without immediate professional care.

The bottleneck caused by insufficient medical staff can be relieved by an AI-powered triage system guiding both trained personnel and civilians for immediate first aid before professional help arrives. Assuming a post-disaster condition, computational constraints can be alleviated to improve emergency response with UAVs equipped with AI-enhanced computing capabilities. Deployed air computing resources process the offloaded real-time tasks, perform data analysis, and provide network extension in affected areas.

\begin{figure*}[!t]
\begin{subfigure}{0.45\textwidth}
\includegraphics[width=\linewidth]{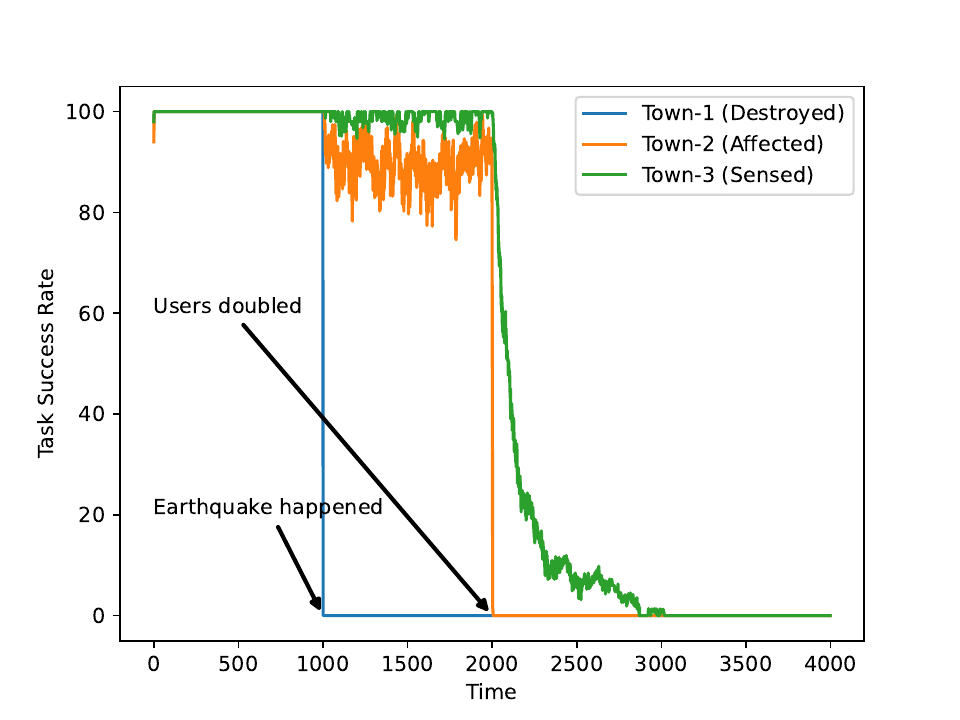}
\caption{Terrestrial infrastructure only} \label{NoUAVTime}
\end{subfigure}
\hspace*{\fill} 
\begin{subfigure}{0.45\textwidth}
\includegraphics[width=\linewidth]{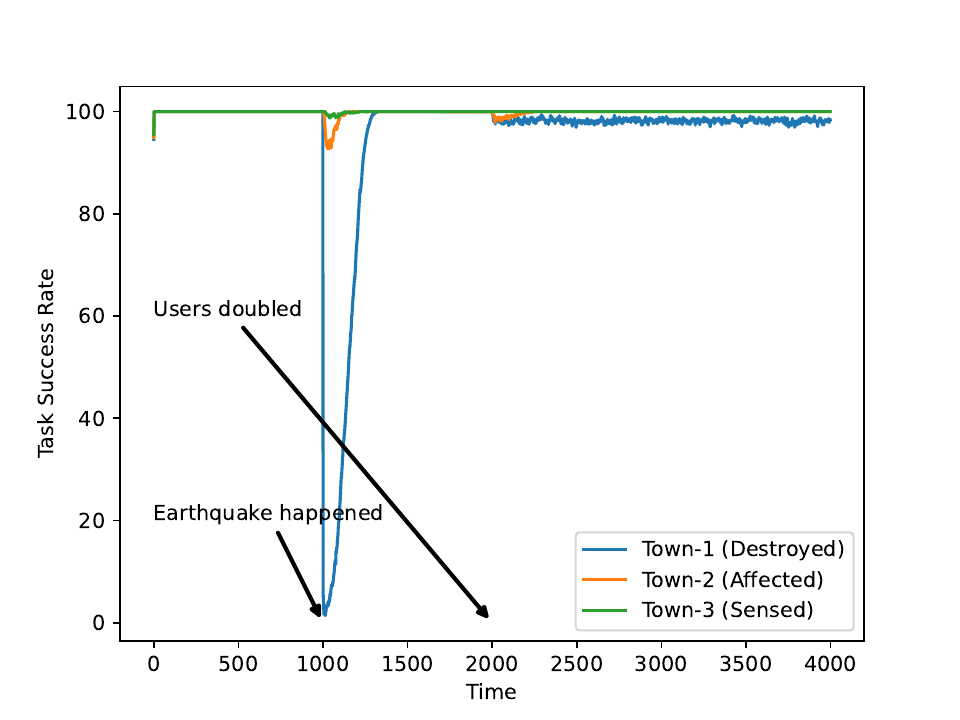}
\caption{Terrestrial infrastructure assisted with air computing} \label{AirComputingLLM}
\end{subfigure}
\caption{Disaster response evaluated with and without air computing support}
\label{OverallTimeResults}
\end{figure*}

\subsection{Parameters}
In order to assess the feasibility of using air computing on disaster areas we conducted a series of simulation experiments. The duration of our simulation is 4000 seconds in which the earthquake happens at 1000 seconds. Before the earthquake, each user in Town-1 and Town-2 uses an application whose required number of CPU units, tolerable delay, and mean interarrival time are 90 units, 1-second, and 3.33 seconds, respectively. These parameters are the same for Town-3 except for the tolerable delay, which is 2 seconds. 

After the earthquake at 1000 seconds, utilization of applications would differ based on the conditions of the towns and the effect on users. Therefore, the utilization of applications triples for each town due to the panic of people so that their new interarrival time is 1 second. Note that the edge server in Town-1 has been destroyed and out of service when the earthquake hit. 

Starting from 2000 seconds, the number of users doubled for each city for several reasons. For Town-1, there are many rescue operations as the town is severely damaged. Moreover, paramedics send the images of injured people along with the explanation to the corresponding multimodal LLMs running on UAVs. On the other hand, those operations are managed from Town-2, as its infrastructure has not been affected. Finally, aftershocks are observed using a facility in Town-3. Note that since rescue operations and their management are critical, the tolerable delay and interarrival time of the new users' tasks are 1 second with 90 CPU units. However, since the observation of aftershocks would not be as critical as those rescue operations, the tolerable delay for new users' tasks in Town-3 is 5 seconds with 12 CPU units. Moreover, their mean interarrival time is 1 second as aftershocks are frequent. 

In simulations, we used edge servers as terrestrial resources. Note that we did not consider cloud servers for offloading since the majority of the tasks would fail considering the wide area network (WAN) delay and required tolerable delay of tasks. Moreover, WAN performance could also be affected by the disaster. Each edge server in each town is identical; an edge server has a capacity of 100K CPU units/sec. We assume that each user in three towns is in the range of an edge server so that they can offload their tasks.

Each UAV in the simulation environment is identical in terms of capacity, radius, and altitude. To this end, a UAV has a capacity of 50K CPU units/sec, a horizontal radius of 100 meters for the offloading range, and an altitude of 200 meters. Note that a UAV does not receive an offloaded task when it is flying towards its destination. Therefore, a task can only be offloaded to a UAV when it arrives at its deployed location. Moreover, all of the offloaded tasks in a UAV queue are processed and returned to the corresponding user regardless of the UAV flying state. 

In our simulations, we consider only the offloaded tasks. If a user is in the range of both an edge server and a UAV, a task is offloaded to the corresponding server whose queueing delay is lower. To this end, we assume that users are informed by the HAP about the queueing conditions of the units. The fixed WLAN delay used in simulations is 1 ms for the tasks offloaded to an edge server, while it is 5 ms for the tasks offloaded to a UAV.

\subsection{Results}

We first observed the outcome of the policy of no UAV case, which can be considered as the baseline. As shown in Figure \ref{NoUAVTime}, the task success rate of Town-1 decreases to 0\% after the earthquake at 1000 seconds since the infrastructure, which is the edge server in our scenario, has been destroyed. Therefore, the corresponding tasks of the users in Town-1 cannot be processed. Moreover, since no UAV is used after the earthquake, the condition of Town-1 does not improve. On the other hand, the task success rate in Town-2 and Town-3 could not be stabilized after the earthquake since the applications are used more frequently so that the capacity of the existing infrastructure cannot be sufficient completely. Especially, when the number of users doubles after 2000 seconds, the task success rate of Town-2 decreases to 0\% regarding the delay requirements of the tasks. Besides, the decline of the task success rate in Town-3 is not as sharp as in Town-2 since the delay requirements of tasks in Town-3 are more tolerant to the queueing delay in the edge server. It is important to note that each town produces the same amount of tasks.

Second, we investigated the effect of air computing and LLM solution on task success rate. To this end, we deployed 8 UAVs by running LLM models on them as a generic task solver. Note that HAP manages the flying policy of these UAVs considering the required capacity for each town. Thus, as shown in Figure \ref{AirComputingLLM}, the task success rate of each town is enhanced dramatically. After the disaster has happened, HAP computes the necessary capacity for each town based on the load of user tasks. Then, it assigns the corresponding UAVs to towns in order to alleviate the existing bottleneck in them. After a couple of seconds of interruption, the task success rate for each town is improved by the system. Thus, QoS and QoE remain satisfactory after the disaster.

\section{Conclusion}

In this study, we investigated the essential features of LLMs and air computing considering their generic task solving abilities, and dynamic capacity enhancement capabilities, respectively. Afterwards, we demonstrated possible use cases in which user experience can be enhanced significantly if they are employed together. In order to manifest the benefits of the collaborative utilization of LLMs and air computing more concretely, we took a disaster scenario into account and conducted experiments. The performance evaluation of those experiments showed that if LLM services are offered over air computing, they alleviate the burden on the terrestrial resources and therefore user tasks would not be affected by the negative outcomes of the disaster. In the future, we plan to evaluate the utilization of air computing and LLMs for different scenarios.

\bibliographystyle{IEEEtran}
\bibliography{LLM-Magazine}

%

\begin{IEEEbiography}
[{\includegraphics[width=1in,height=1.25in,clip,keepaspectratio]{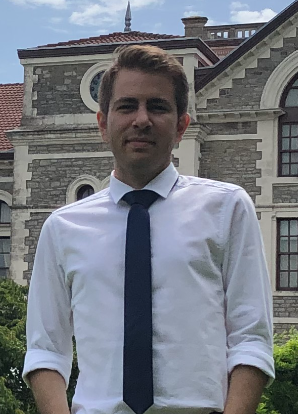}}]{Baris Yamansavacilar}
received his BS degree in Computer Engineering from Yildiz Technical University, Istanbul, in 2015. He received his MS degree in Computer Engineering from Bogazici University, Istanbul, in 2019. Currently, he is a PhD candidate in Computer Engineering Department at Bogazici University and a senior research engineer at Airties. His research interests include Machine Learning, Deep Reinforcement Learning, Edge/Cloud Computing, and Software-Defined Networks.

\end{IEEEbiography}
\begin{IEEEbiography}[{\includegraphics[width=1in,height=1.25in,clip,keepaspectratio]{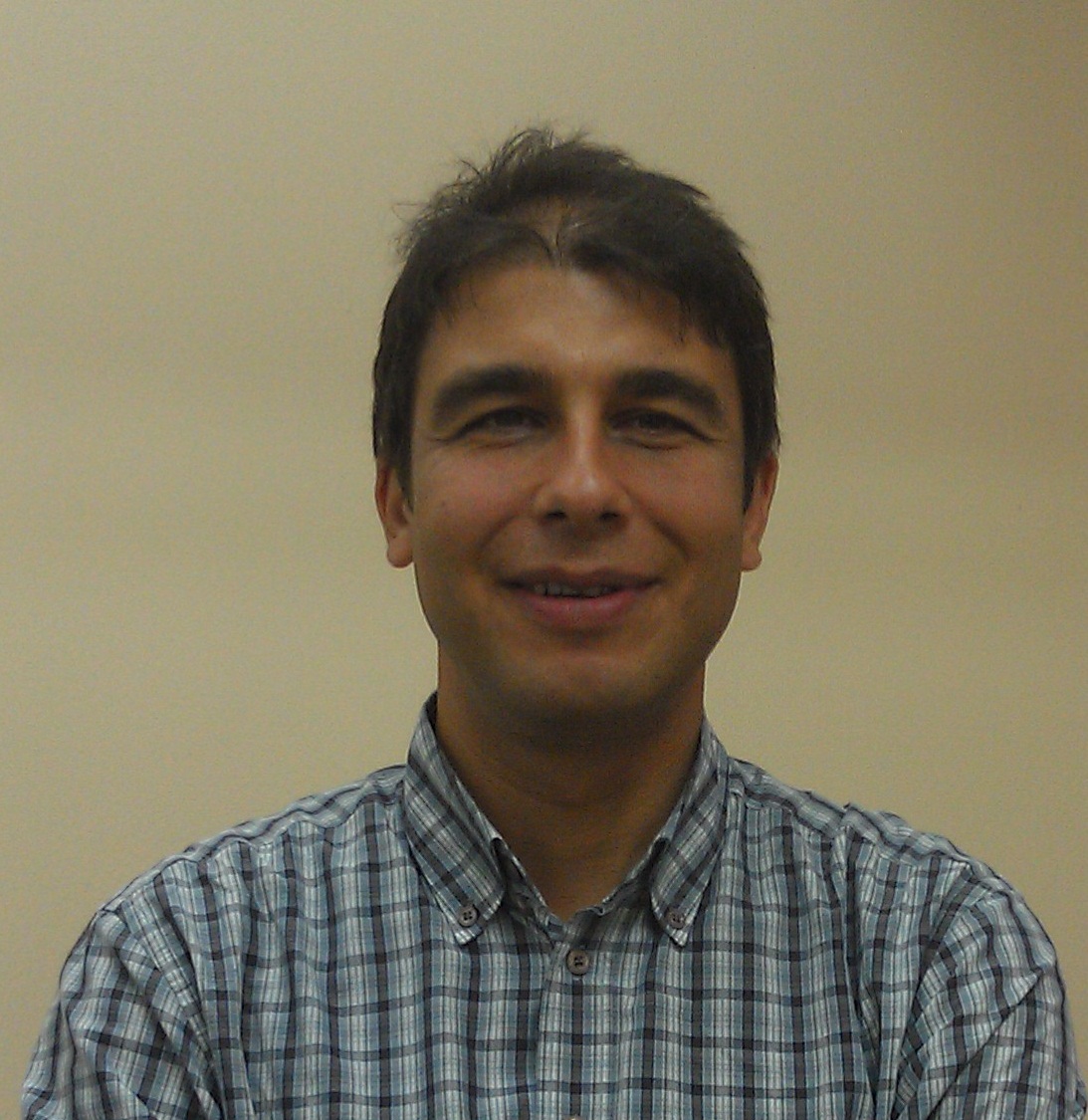}}]{Atay Ozgovde}
received the BS and MS degrees from Bogazici University, Istanbul, in 1995 and 1998, respectively. In 2002, he started working as a research assistant in the Computer Engineering Department, Bogazici University, where he completed the PhD degree in the NETLAB research group in 2009. He worked as a researcher at the WiSE-Ambient Intelligence Group to complete his postdoctoral research. Currently, he is an assistant professor in the Computer Engineering Department, Bogazici University. His research interests include wireless sensor networks, embedded systems, distributed systems, pervasive computing, SDN and mobile cloud computing. He is a member of the IEEE.
\end{IEEEbiography}

\begin{IEEEbiography}[{\includegraphics[width=1in,height=1.25in,clip,keepaspectratio]{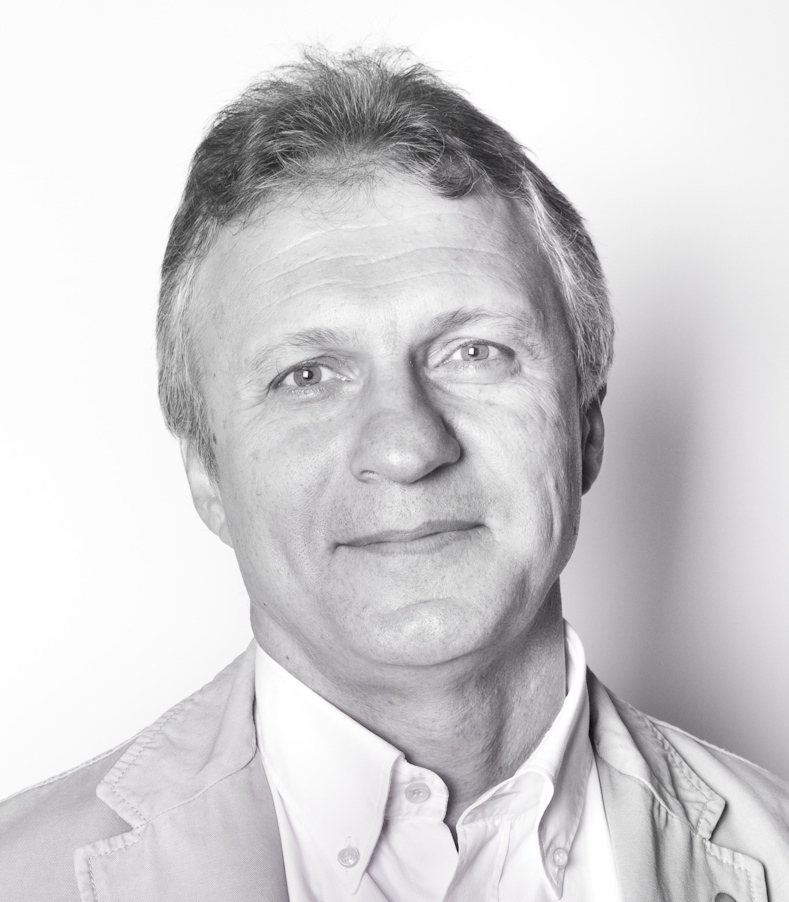}}]{Cem Ersoy}	
worked as an R\&D engineer in NETAS A.S. between 1984 and 1986. After receiving his PhD from Polytechnic University, New York in 1992, he became a professor of Computer Engineering in Bogazici University. Prof. Ersoy's research interests include 5G and beyond networks, mobile cloud/edge/fog computing. Prof. Ersoy is a member of IFIP and was the chairperson of the IEEE Communications Society Turkish Chapter for eight years.
\end{IEEEbiography}

\vfill

\end{document}